\documentclass[%
 aip,
 amsmath,amssymb,
 reprint,%
]{revtex4-1}

\usepackage{graphicx}
\usepackage{dcolumn}
\usepackage{bm}

\usepackage[utf8]{inputenc}
\usepackage[T1]{fontenc}
\usepackage{mathptmx}

\usepackage{cleveref}
\usepackage{caption}
\begin{document}

\preprint{AIP/123-QED}

\title[Tight-Binding Bandstructure of $\beta-$ and $\alpha-$ phase Ga$_2$O$_3$ and Al$_2$O$_3$]{Tight-Binding Bandstructure of $\beta-$ and $\alpha-$ phase Ga$_2$O$_3$ and Al$_2$O$_3$}

\author{Y. Zhang}
\email{yz2439@cornell.edu.}
\affiliation{Electrical and Computer Engineering, Cornell University, Ithaca, NY 14850, USA
}%

\author{M. Liu}
\affiliation{Materials Science and Engineering, Cornell University, Ithaca, NY 14850, USA
}%


\author{G. Khalsa}
\email{guru.khalsa@cornell.edu.}
\affiliation{Materials Science and Engineering, Cornell University, Ithaca, NY 14850, USA
}%

\author{D. Jena}
\affiliation{Electrical and Computer Engineering, Cornell University, Ithaca, NY 14850, USA
}%
\affiliation{Materials Science and Engineering, Cornell University, Ithaca, NY 14850, USA
}%

 


\date{\today}

\begin{abstract}
Rapid design and development of the emergent ultra-wide bandgap semiconductors Ga$_2$O$_3$ and Al$_2$O$_3$ requires a compact model of their electronic structures, accurate over the broad energy range accessed in future high-field, high-frequency, and high-temperature electronics and visible and ultraviolet photonics. A minimal tight-binding model is developed to reproduce the first-principles electronic structures of the $\beta-$ and $\alpha-$ phases of Ga$_2$O$_3$ and Al$_2$O$_3$ throughout their reciprocal spaces. Accurately reproducing the bandgap, orbital character, and effective mass and high-energy features of the conduction band, this compact model will assist in the investigation and design of the electrical and optical properties of bulk materials, devices, and quantum confined heterostructures.
\end{abstract}

\maketitle

\section{Introduction}

The recent integration of Ga$_2$O$_3$ with Al$_2$O$_3$ has the potential to revolutionize high-power electronics. The availability of large, inexpensive, single-crystal substrates \cite{Higashiwaki2012, Kuramata2016}, recent advances in thin film growth\cite{Gogova2014,Okumura2014,Jinno2021}, and the ability to dope these wide-bandgap semiconductors have enabled transistors and Schottky diodes based on $\beta$-Ga$_2$O$_3$ with breakdown fields as large as 5.45 MV/cm (Ref. \onlinecite{Roy2021}) and 5.7 MV/cm (Ref. \onlinecite{Kalarickal2021}) and approaching the projected theoretical estimate of 8 MV/cm (Ref. \onlinecite{Pearton2018}). Comparing these breakdown fields with the existing technological semiconductors Si (0.3 MV/cm), SiC (3.1 MV/cm), and GaN (3.3 MV/cm) \cite{Meneghini2017}, $\beta$-Ga$_2$O$_3$ promises new high-frequency, high-voltage, and high-temperature electronics applications. $\alpha$-Ga$_2$O$_3$ and $\alpha$-Al$_2$O$_3$ further expand the bandgap to 5.2 eV and 8.8 eV, signifying the potential for oxide semiconductors to expand the future electronics materials tool-set.

The successful design of these future electronic devices requires an accurate modeling and understanding of the electronic structure and bonding of Ga$_2$O$_3$ and Al$_2$O$_3$. The tight-binding method provides a flexible, chemically motivated description of the electronic structure of materials\cite{electronic-structure}. When compared with modern computational approaches to materials physics like density functional theory (DFT), tight-binding models are compact, intuitive, and require less computational resources. As a result, tight-binding models are ubiquitous in device engineering and development and have successfully described electronic transport\cite{datta_2005,Lake1997, Nehari2005,Popov2004,Sols1989} and optical properties\cite{Graf1995,Popov2004,Pi2013,Jirauschek2014} of bulk materials, heterostructures, and devices. To aid in the development of new high-power electronics, we derive semi-empirical tight-binding models in this work for three technologically relevant oxide semiconductors: $\beta$-Ga$_2$O$_3$, $\alpha$-Ga$_2$O$_3$, and $\alpha$-Al$_2$O$_3$. 

While we are unaware of a tight-binding model describing these three oxide semiconductors, a recent study reports a tight-binding model of $\beta$-Ga$_2$O$_3$ using atomic orbitals as a basis, with parameters drawn from DFT calculations\cite{Lee2019}. The authors employ the model to study the surface energy of $\beta$-Ga$_2$O$_3$ and formation energy of Ga and O vacancy defects. We derive an alternative tight-binding model with the goal of accurate parameterization of the conduction band and fundamental optical gaps of $\beta$-Ga$_2$O$_3$, $\alpha$-Ga$_2$O$_3$ and $\alpha$-Al$_2$O$_3$ so that electrical and optical properties can be faithfully simulated.

We derive tight-binding models using a Wannier functions basis. Wannier functions are a convenient basis for tight-binding models because they are derived from the underlying band structure of the material, are formally orthogonal, can be localized to atomic sites, and preserve the site symmetry and coordination. This approach of DFT-derived tight-binding has been used successfully to describe the electronic structure of broad classes of technologically important materials including silicon\cite{Souza2002}, III-V semiconductors\cite{Gresch2018}, and 2D materials\cite{Souza2012}.

\section{Structural and Electrical Properties}

\begin{figure*}
\scalebox{.4}{\includegraphics{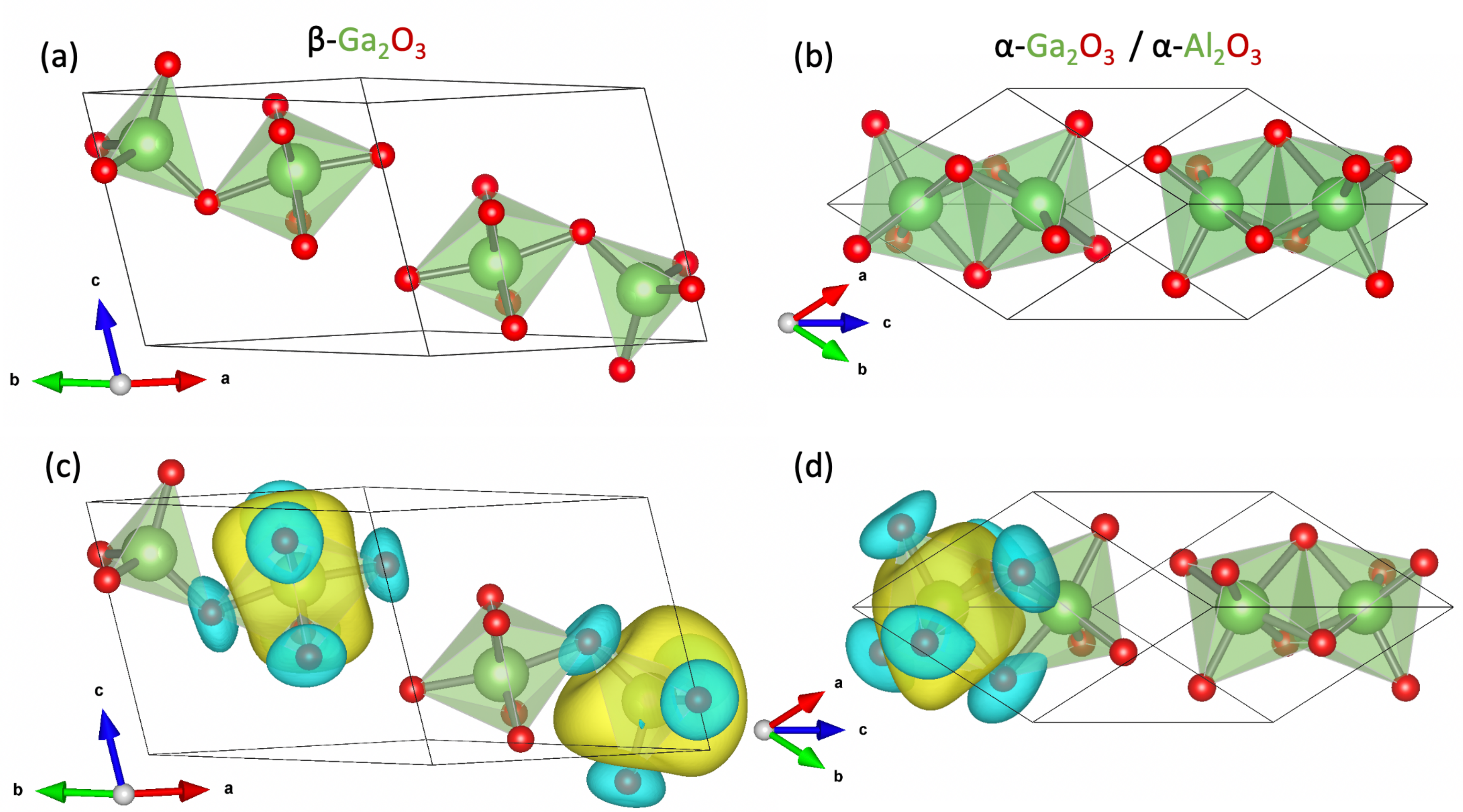}}
\caption{\label{fig:crystal} 
Crystal structure of monoclinic $\beta$-Ga$_2$O$_3$, and rhombohedral $\alpha$-Ga$_2$O$_3$ and $\alpha$-Al$_2$O$_3$. Coordination octahedra and tetraheda highlight the different Ga(Al)-O bonding environments in the $\beta$ (a) and $\alpha$ (b) phases. This bonding difference is evident in the symmetry of Ga(Al)-site Wannier functions (c,d) for the conduction band where clear hybridization of the Ga(Al)-s and O-p is seen. In (c,d), the positive lobes of the Wannier functions are shown in yellow and the negative lobes are shown in blue.}
\end{figure*}

The crystal symmetry and bonding environment constrain the tight-binding description of the electronic structure. When compared to a conventional semiconductor like Si, Ga$_2$O$_3$ and Al$_2$O$_3$ have relatively low symmetry and complicated bonding networks. $\beta$-Ga$_2$O$_3$ has a monoclinic structure (space group C2/m, No. 12). The monoclinic structure contains two pairs of symmetry inequivalent Ga sites, each coordinated by O, forming two distorted GaO$_4$ tetrahedra and two distorted GaO$_6$ octahedra per unit cell (Fig.~\ref{fig:crystal}(a)). $\alpha$-Al$_2$O$_3$ and $\alpha$-Ga$_2$O$_3$ crystallize in the sapphire structure (rhombohedral, space group R-3c, No. 167). In the $\alpha$ phase, the Al(Ga) atoms occupy 4 equivalent sites, each coordinated by six O, forming distorted AlO$_6$(GaO$_6$) octahedra (Fig.~\ref{fig:crystal}(b)).  The structural information obtained from DFT structural optimization and experimental data are given in \Cref{tab:struct_comp} where DFT is shown to describe the experimental structure surprisingly well. The valence configurations of O, Al, and Ga are 2s$^2$2p$^4$, 3s$^2$3p$^1$, and 4s$^2$4p$^1$, respectively. The Al(Ga) is expected to donate its valence electrons in order to fill the O valence shell, leading to an O-2p derived valence band with Al-3s(Ga-4s) and Al-3p(Ga-4p) derived conduction band. 

Structural optimization and electronic band structure are calculated by DFT using Quantum Espresso\cite{Giannozzi_2009}. We choose projector-augmented wave pseudopotentials and generalized-gradient approximation of exchange correlation using the Perdew-Burke-Ernzerhof functional generalized for solids\cite{Perdew1996,Perdew2008}. Structural convergence is found for a $3 \times 3 \times 3$ k-point mesh and a 60 Ry plane-wave cutoff. We find good agreement between the DFT-relaxed structure and the experimental structure (see \Cref{tab:struct_comp}). 

\begin{table*}
\caption{\label{tab:struct_comp} $\beta$-Ga$_2$O$_3$, $\alpha$-Ga$_2$O$_3$, and $\alpha$-Al$_2$O$_3$ structural data. In the $\beta$-phase (left column), Ga and O occupy the $i$ Wyckoff site with fractional coordinates $(x,0,z)$ and $(-x,0,-z)$. In the $\alpha$ phase (right column), Ga(Al) occupy the $c$ Wyckoff site with fractional coordinates $(z,z,z)$, 	$(-z+\frac{1}{2},-z+\frac{1}{2},-z+\frac{1}{2})$, $(-z,-z,-z)$, and $(z+\frac{1}{2},z+\frac{1}{2},z+\frac{1}{2})$ and $O$ occupies the $e$ Wyckoff site with fractional coordinates $(x+\frac{1}{4},-x+\frac{1}{4},\frac{1}{4})$, $(\frac{1}{4},x+\frac{1}{4},-x+\frac{1}{4})$, $(-x+\frac{1}{4},\frac{1}{4},x+\frac{1}{4})$, $(-x+\frac{3}{4},x+\frac{3}{4},\frac{3}{4})$, $(\frac{3}{4},-x+\frac{3}{4},x+\frac{3}{4})$, and $(x+\frac{3}{4},\frac{3}{4},-x+\frac{3}{4})$.}
\begin{minipage}{0.5\linewidth}
\begin{tabular}{ccccc}
\hline
\hline
$\beta$-Ga$_2$O$_3$ & $a$ (\AA) & $b$ (\AA) & $c$ (\AA) & $\beta$ ($^\circ$) \\
\hline
DFT & 12.237 & 3.062 & 5.813 & 103.81\\
Experiment\cite{Zade2019} & 12.214 & 3.037 & 5.798 & 103.83\\
& & & & \\
\end{tabular}
\begin{tabular}{ccccc}
\hline
 \multicolumn{1}{c|}{}& \multicolumn{2}{c|}{Ga(I) (4i)} & \multicolumn{2}{c}{Ga(II) (4i)} \\
\multicolumn{1}{c|}{$\beta$-Ga$_2$O$_3$} & $x$ & \multicolumn{1}{c|}{$z$}      &  $x$  & $z$      \\
\hline
\multicolumn{1}{c|}{DFT} & 0.0904 & \multicolumn{1}{c|}{0.7949} & 0.6585 & 0.3137  \\
\multicolumn{1}{c|}{Experiment\cite{Zade2019}} & 0.0895 & \multicolumn{1}{c|}{0.7938} & 0.6585 & 0.3100 \\
& & \\
\end{tabular}
\begin{tabular}{ccccccc}
\hline
 \multicolumn{1}{c|}{}& \multicolumn{2}{c|}{O(I) (4i)} & \multicolumn{2}{c|}{O(II) (4i)} & \multicolumn{2}{c}{O(III) (4i)} \\
\multicolumn{1}{c|}{$\beta$-Ga$_2$O$_3$} & $x$ & \multicolumn{1}{c|}{$z$} & $x$ & \multicolumn{1}{c|}{$z$} &  $x$  & $z$      \\
\hline
\multicolumn{1}{c|}{DFT} & 0.1647 & \multicolumn{1}{c|}{0.1094} & 0.1733 & \multicolumn{1}{c|}{0.5632} & 0.4959 & 0.2563  \\
\multicolumn{1}{c|}{Experiment\cite{Zade2019}} & 0.1519 & \multicolumn{1}{c|}{0.1001} & 0.1722 & \multicolumn{1}{c|}{0.5640} & 0.4920 & 0.2645 \\
\end{tabular}
\end{minipage}%
\begin{minipage}{.5\linewidth}
\begin{tabular}{ccccc}
\hline
\hline
$\alpha$-Ga$_2$O$_3$ & $a$ (\AA) & $b$ (\AA) & Ga(I) 4c & O(I) 6e \\
\hline
DFT & 5.3221 & 55.82 & $z$=0.1446 & $x$=0.3049\\
Experiment\cite{Marezio1967} & 5.3221 & 55.82 & $z$=0.1446 & $x$=0.3049\\
\end{tabular}
\begin{tabular}{ccccc}
 & & & & \\
 & & & & \\
\hline
\hline
$\alpha$-Al$_2$O$_3$ & $a$ (\AA) & $b$ (\AA) & Al(I) 4c & O(I) 6e \\
\hline
DFT & 5.1779 & 55.28 & $z$=0.1479 & $x$=0.3056 \\
Experiment\cite{Ishizawa1980} & 5.126 & 55.25 & $z$=0.3522 & $x$=0.3064 \\
\end{tabular}
\end{minipage}%
\end{table*}

\begin{figure*}
\scalebox{.4}{\includegraphics{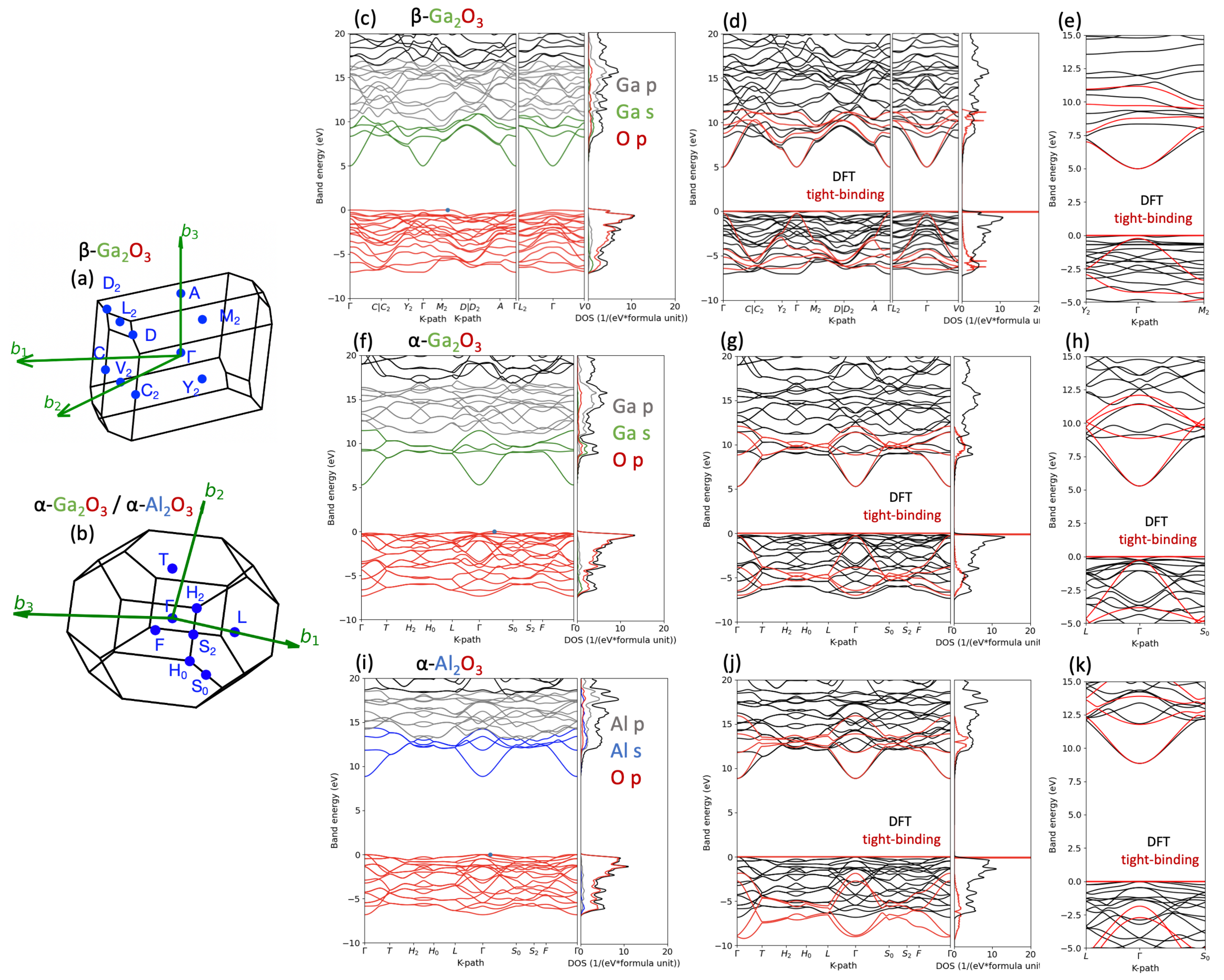}}
\caption{\label{fig:band}
Electronic structure of $\beta$-Ga$_2$O$_3$, $\alpha$-Ga$_2$O$_3$, and $\alpha$-Al$_2$O$_3$. (a,b) -- The first Brillouin zone of the monoclinic $\beta$ and rhombohedral $\alpha$ phases. (c,f,i) -- The DFT band structure and orbital-projected DOS of $\beta$-Ga$_2$O$_3$, $\alpha$-Ga$_2$O$_3$, and $\alpha$-Al$_2$O$_3$. The DFT bandgaps have been tuned to the experimental bandgaps via a scissor cut. Color coordination indicates the orbital character of the bands and projected DOS. The blue dots indicate the valence band maximum. (d,g,j) -- The tight-binding band structure and DOS (red) plotted over the DFT data (black). The sharp peak in the tight-binding DOS at the top of the valence band is due to the lack of O-p to O-p coupling in the tight-binding model.
(e,h,k) -- DFT (black) and tight-binding (red) band structure near the $\Gamma$ point. The reciprocal space vectors and high-symmetry points are given in Table SII.
}
\end{figure*}

The DFT band structure and DOS are shown in Fig.~\ref{fig:band}(c,f,i) and found to be in agreement with previous work\cite{Peelaers2015a}. Orbital projected DOS shows the valence band is primarily O-2p and the conduction band is primarily Ga-4s(Al-3s) and Ga-4p(Al-3p). The bottom of the conduction band is dominated by Ga-4s(Al-3s) states, transitioning to mainly Ga-4p(Al-3p) in character at 6 eV above the conduction band minimum. All three materials share several broad features of their band structure: broad O-2p valence bands with a flat valence band edges, conduction band edge features near the $\Gamma$ point, and indirect bandgaps. In all three materials, the valence band edge is populated by flat O-2p bands. DFT predicts the valence band maximum between M$_2$ and D in $\beta$-Ga$_2$O$_3$, and between $\Gamma$ and S$_0$ in both $\alpha$-Ga$_2$O$_3$ and $\alpha$-Al$_2$O$_3$. The conduction band minimum at the $\Gamma$-point has nearly isotropic dispersion, suggesting that it is insensitive to symmetry and chemistry. We find an effective mass of 0.26 $m_e$ in $\beta$-Ga$_2$O$_3$, 0.28 $m_e$ in $\alpha$-Ga$_2$O$_3$, and 0.42 $m_e$ in $\alpha$-Al$_2$O$_3$. We found that the conduction band effective mass of $\beta$-Ga$_2$O$_3$ is in good agreement with recent transport\cite{Zhang2018} and angle-resolved photoemission spectroscopy\cite{Mohamed2010,Janowitz2011} (ARPES) experiments. Away from the band minimum, the conduction band transitions from parabolic to linear dispersion, a band structure feature used to explain high field transport\cite{Ghosh2017,Ghosh2018} and optical absorption\cite{Singh2020} experiments. The second conduction band at the $\Gamma$ point is above the conduction band minimum by 3.3 eV in $\beta$-Ga$_2$O$_3$, 3.5 eV in $\alpha$-Ga$_2$O$_3$, and 3.1 eV in $\alpha$-Al$_2$O$_3$, respectively. The value of $\beta$-Ga$_2$O$_3$ agrees with the experimental observation of 3.55 eV\cite{Singh2020}. DFT predicts an indirect bandgap with the difference between the direct and indirect bandgap less than 20 meV. This is confirmed by the ARPES measurement\cite{Mohamed2010,Janowitz2011}. As expected, the bandgap predicted by DFT underestimates the experimental bandgaps. In order to match the experimental bandgap, we have conducted a ``scissor cut'' by shifting the conduction band states up in energy so that the bandgaps match the experimental values of 4.9 eV\cite{Janowitz2011}, 5.2 eV\cite{Shinohara2008}, and 8.8 eV\cite{French1990} in $\beta$-Ga$_2$O$_3$, $\alpha$-Ga$_2$O$_3$, and $\alpha$-Al$_2$O$_3$, respectively. In constructing a tight-binding model, we aim to describe the above key features of the DFT band structure.

\section{Tight-Binding Derivation}

We derive Wannier functions from the DFT band structure using Wannier90\cite{Pizzi_2020} to construct the tight-binding basis. Wannier functions are initialized on the Ga(Al) sites with s- and p-orbital symmetry and on the O sites with p-orbital symmetry. We selectively localize only the Ga-s(Al-s) Wannier functions onto Ga(Al) sites using a Lagrange multiplier\cite{Wang2014}. Selectively-localized Wannier functions improve the localization of Ga-s(Al-s) Wannier functions at the cost of delocalizing Ga-p(Al-p) Wannier functions which describe the higher conduction bands (> 5 eV above the conduction band minimum). We find that the O-p Wannier functions have similar localization in both schemes. 

We show isosurface plots of Ga-s Wannier functions in $\beta$ and $\alpha$ phases resulting from selective localization in Fig.~\ref{fig:crystal}(c,d). (The Al-s and Ga-s Wannier functions in the $\alpha$ phase are qualitatively similar.) The Wannier functions reproduce the distorted tetrahedral and octahedral symmetry of the coordination polyhedra as expected from the hybridization between the Ga-4s and O-2p atomic orbitals in Ga$_2$O$_3$. 

We construct the tight-binding model from the Wannier function basis by extracting the Hamiltonian matrix element using PythTB\cite{Coh2013}. We parameterize our tight-binding model in the following form:

\begin{equation}
\label{eq:one}
\begin{split}
 \hat{H}&=\sum_i\epsilon_i\left. |i\right>\left<i|\right.\\
 &+\sum_{i,j}t_{ij}e^{-i\vec{k}\cdot\vec{\Delta d_{ij}}}\left. |j\right>\left<i|\right.+c.c.
 \end{split}
\end{equation}

\noindent
The right-hand side of the first line describes the contribution of individual Wannier functions to the total energy, commonly called the onsite energy. The second line describes the kinetic energy of the electrons, commonly called the ``hopping" energy. In \Cref{eq:one}, $\left. |i\right>$ represents the $i^{th}$ Wannier function with on-site energy $\epsilon_i$.  $t_{ij}=\tilde{t}_{ij}e^{i\phi_{ij}}$ is the hopping term between $i^{th}$ and $j^{th}$ Wannier functions. Here $\phi_{ij}$ characterizes the phase of $t_{ij}$ and encompasses the symmetry of the local bonding environment.
$\vec{k}$ is the wavevector, and $\Delta \vec{d}_{ij}$ is the displacement vector of the two Wannier functions $i$ and $j$ defined as 
$\vec{\Delta d_{ij}}=\vec{d_j}-\vec{d_i}+\vec{R}$.
Here $\vec{d}_i$ and $\vec{d}_j$ are the centers of Wannier function $i$ and $j$ within the same unit cell, and $\vec{R}$ is the unit cell translation to indicate coupling across adjacent unit cells.

In our tight-binding model, we include only the s-orbital-derived Wannier functions from the 4 Ga(Al) atoms and the 3 p-orbital-derived Wannier functions from the 6 O atoms. This translates to a {22 $\times$ 22} matrix when the model is implemented numerically. To describe the band structure with a minimal set of parameters, we include: 1) the onsite energy ($\epsilon_{i}$ terms for Ga-s(Al-s) and O-p, 2) the nearest-neighbor Ga-s(Al-s) to O-p and, 3) the dominant Ga-s(Al-s) to Ga-s(Al-s) terms ($t_{ij}$). This amounts to a nearest-neighbor tight-binding model augmented by the Ga-s(Al-s) to Ga-s(Al-s) next-nearest-neighbor hopping. These next-nearest-neighbor terms aid in the accuracy of the higher Ga-4s(Al-3s) derived conduction bands. With these simplifications, the model contains ~60 parameters depending on the structural phases (see Table SIII to SV). Including the Ga-p(Al-p) Wannier functions and O-p to O-p coupling terms gives a satisfactory description of valence band DOS at the cost of significantly increasing the number of parameters (around 300 terms), but provides little improvement to the description of the lower conduction bands. Thus, we neglect these terms in our model. 

To reproduce the experimental bandgaps and conduction band effective masses, we adjust the Ga-s(Al-s) onsite energy and tune the Ga-s(Al-s) to O-p coupling term. (We note that in their model of $\beta$-Ga$_2$O$_3$, Lee, et al.\cite{Lee2019} focus on surface and defect formation energies which depend on the broad features of the band structure. As a result, their tight-binding model overestimates the conduction band effective mass.) The lists of parameters are given in Table SIII, SIV, and SV.

When implementing the tight-binding model on a computer, Eq.~(\ref{eq:one}) can be written as a 22 $\times$ 22 matrix defined over the basis of 4 Ga-s(Al-s) and 18 O-p Wannier functions. In evaluating the model, the on-site energy  $\epsilon_i$ become diagonal terms while the hopping terms $t_{ij}$ populate the i$^{th}$ row and j$^{th}$ column and must be multiplied by the phase factor $e^{-i\vec{k}\cdot\vec{\Delta d_{ij}}}$ at each k-point. To illustrate the model and gain insight into the material physics, we explicitly evaluate the tight-binding model of $\beta$-Ga$_2$O$_3$ at the $\Gamma$-point ($\vec{k}=\vec{0}$). We find that the tight-binding model at the $\Gamma$-point can be written in the block-matrix form:

\begin{equation}
    H=\begin{pmatrix}
        H_{Ga:s} & H_{Ga:s,O:p_x}^\dagger & 0 & H_{Ga:s,O:p_z}^\dagger \\
        H_{Ga:s,O:p_x} & H_{O:p_x} & 0 & 0 \\
        0 & 0 & H_{O:p_y} & 0 \\
        H_{Ga:s,O:p_z} & 0 & 0 & H_{O:p_z}
\end{pmatrix}
\end{equation}

\noindent
where the $H_{Ga:s}$, $H_{O:p_x}$, $H_{O:p_y}$, and $H_{O:p_z}$ blocks define the coupling within the Ga-s, O-p$_x$, O-p$_y$, and O-p$_z$ Wannier function sub-spaces, respectively. The $H_{Ga:s,O:p_x}$ and $H_{Ga:s,O:p_z}$ blocks describe the coupling between the Ga-s and O-p$_x$, and O-p$_z$ Wannier functions, respectively. We construct the tight-binding model so that it reflects the crystal symmetry. This can be seen clearly in the lack of coupling between the Ga-s and O-p$_y$ Wannier functions. The two-fold rotation about the crystallographic b-axis and the mirror operation through the plane perpendicular to the b-axis guarantee this coupling is zero at the $\Gamma$-point. The coupling between different O-p Wannier function blocks (e.g. O-p$_x$ and O-p$_y$) are zero because we have neglected the coupling within the valence band. At the $\Gamma$ point, the phase factor $e^{-i\vec{k}\cdot\vec{\Delta d_{ij}}} \rightarrow 1$, leaving the matrix real and symmetric.

$H_{Ga:s}$ takes the form:

\begin{widetext}
\begin{equation}
\begin{split}
H_{Ga:s} &=\begin{pmatrix}
        \epsilon_{Ga1:s} + 2t_{Ga1:s,Ga1:s}& 0 & 0 & 0\\
        0 & \epsilon_{Ga1:s} +2t_{Ga1:s,Ga1:s} & 0 & 0 \\
        0 & 0 & \epsilon_{Ga3:s} + 2t_{Ga3:s,Ga3:s} & t_{Ga3:s,Ga4:s}  \\
        0 & 0 & t_{Ga3:s,Ga4:s} & \epsilon_{Ga3:s} + 2t_{Ga3:s,Ga3:s} 
\end{pmatrix}\\
&\rightarrow
\begin{pmatrix}
        -7.498 & 0 & 0 & 0\\
        0 & -7.498 & 0 & 0 \\
        0 & 0 & -7.222 & 0.216 \\
        0 & 0 & 0.216 & -7.222
\end{pmatrix}
\end{split}
\end{equation}
\end{widetext}
\noindent
which is written in the basis $\left(\left| Ga1:s \right>,\left| Ga2:s \right>,\left| Ga3:s \right>,\left| Ga4:s \right>\right)$ of Ga-s Wannier functions. Taking values from Table SIII, we have included the numerical value of the block in eV. Since Ga1:s and Ga2:s are equivalent tetrahedral sites while Ga3:s and Ga4:s are equivalent octahedral sites, each have the same onsite energy and coupling. Notice that the onsite energy $\epsilon_{Ga:1s}$ and $\epsilon_{Ga:3s}$ are modified by coupling to the same Ga sites in the neighboring unit cells ($t_{Ga1:s,Ga1:s}$ and $t_{Ga3:s,Ga3:s}$). Furthermore, the octahedral sites are weakly coupled to each other. 

In simplifying the description of the valence band, we neglect the coupling between O-p Wannier functions. As a result each $H_{O:px}$, $H_{O:py}$, and $H_{O:pz}$ block is diagonal and can be written as:

\begin{equation}
H_{O:px}=H_{O:py}=H_{O:pz}=\epsilon_{O:p}\times \mathbb{I}_{6\times 6}\rightarrow-12\times \mathbb{I}_{6\times 6}
\end{equation}

\noindent which is written in the basis $\left(\right. \left| O1:p_l \right>$,$\left| O2:p_l\right>$,$\left| O3:p_l \right>$,$\left| O4:p_l \right>$,$\left| O5:p_l \right>$,$\left| O6:p_l \right>\left.\right)$, of O-p Wannier functions where $l=x,y,z$ . Here, $\mathbb{I}_{6\times6}$ is the 6 $\times$ 6 identity matrix.

The coupling between Ga-s and O-p$_x$ and O-p$_z$ are described by $H_{Ga:s,O:px}$ and $H_{Ga:s,O:pz}$ which evaluate to (hopping strengths connected by symmetries are shown with the same symbol):

\begin{widetext}
\begin{equation}
H_{Ga:s,O:px}=
\begin{pmatrix}
        t_{Ga1:s,O1:px} & 0 & 2t_{Ga3:s,O1:px} & 0\\
        0 & -t_{Ga1:s,O1:px} & 0 & 2t_{Ga4:s,O2:px} \\
        t_{Ga1:s,O3:px} & 0 & 0 & -t_{Ga3:s,O4:px} \\
        0 & t_{Ga2:s,O4:px} & t_{Ga3:s,O4:px} & 0 \\
        0 & -2t_{Ga1:s,O6:px} & t_{Ga3:s,O5:px} & 0 \\
        2t_{Ga1:s,O6:px} & 0 & 0 & -t_{Ga3:s,O5:px} \\
\end{pmatrix}
\rightarrow\begin{pmatrix}
        0.651 & 0 & 0.710 & 0\\
        0 & -0.651 & 0 & -0.703 \\
        2.592 & 0 & 0 & -2.877 \\
        0 & -2.595 & 2.877 & 0 \\
        0 & 3.499 & -2.965 & 0 \\
        -3.496 & 0 & 0 & 2.965 \\
\end{pmatrix}
\end{equation}
\begin{equation}
H_{Ga:s,O:pz}=
\begin{pmatrix}
        t_{Ga1:s,O1:pz} & 0 & 2t_{Ga3:s,O1:pz} & 0\\
        0 & -t_{Ga1:s,O1:pz} & 0 & 2t_{Ga4:s,O2:pz} \\
        t_{Ga1:s,O3:px} & 0 & 2t_{Ga3:s,O3:pz} & -t_{Ga3:s,O4:pz} \\
        0 & -t_{Ga2:s,O4:px} & t_{Ga3:s,O4:pz} & -2t_{Ga4:s,O4:pz} \\
        0 & -2t_{Ga1:s,O6:pz} & t_{Ga3:s,O5:pz} & 0 \\
        2t_{Ga1:s,O6:pz} & 0 & 0 & -t_{Ga3:s,O5:pz} \\
\end{pmatrix}
\rightarrow\begin{pmatrix}
        3.466 & 0 & -3.307 & 0\\
        0 & -3.466 & 0 & 3.314 \\
        2.592 & 0 & 3.345 & -1.101 \\
        0 & 2.595 & 1.101 & -3.326 \\
        0 & 0.885 & -0.621 & 0 \\
        -0.880 & 0 & 0 & 0.621 \\
\end{pmatrix}
\end{equation}
\end{widetext}

\noindent Here, terms appear as pairs with opposite signs, manifesting the two-fold symmetry of the monoclinic structure. $H_{Ga:s,O:px}$ and $H_{Ga:s,O:pz}$ looks formally similar but are not exactly the same. Again, the matrix coupling Ga-s to O-py vanishes at the $\Gamma$ point but will show up away from it. The complete set of matrix elements can be found in Table SIII. The eigenvalues of the 22 $\times$ 22 matrix correspond to the energy eigenstates at the given k-point, and the eigenvectors correspond to the wavefunctions. To generate the band structure, one generates the matrix at sampled k-points along the path and solve for the eigenvalues. The complexity of numerical eigensolvers depends on the matrix dimension as $O(n^3)$. As a comparison, our DFT calculation relies on around 5000 planar waves as a basis, so the tight-binding model provides a significant speedup. In our experience, a tight-binding band structure can be generated on a personal computer in seconds.

\section{Tight-Binding Bandstructure}

Fig.~\ref{fig:band}(d,g,j) shows the tight-binding band structure and DOS superimposed on the DFT results. In all three phases, the experimental bandgaps and conduction band effective masses are reproduced by the adjustment of parameters mentioned above. Moreover, the tight-binding model gives a satisfactory description of the parabolic to linear dispersion, the second conduction band at the $\Gamma$ point, and the flat valence band edge states.

The slope of the linear dispersion of the lowest conduction band away from $\Gamma$, and the energy of the second conduction band at $\Gamma$ are slightly overestimated due to the absence of interaction from higher Ga-p(Al-p) bands. This phenomenon is well-known in the tight-binding description of Si, Ge, and III-V semiconductors\cite{Vogl1983} where it is due to the lack of interactions from higher energy bands. The lack of O-p to O-p coupling leaves some of the O-p states non-dispersive. As a result, the tight-binding model cannot describe the small difference between the valence band at $\Gamma$ and the DFT predicted valence band maximum between $\Gamma$ and T in the $\beta$ phase and $\Gamma$ and $S_0$ in the $\alpha$ phases. This leaves a large peak in the DOS at the valence band edge (see Fig.~\ref{fig:band}(d-f)) which could be resolved in future models that include O-p to O-p coupling.

\section{Conclusion}

We have derived a minimal tight-binding model for $\beta$-Ga$_2$O$_3$, $\alpha$-Ga$_2$O$_3$, and $\alpha$-Al$_2$O$_3$ using selectively-localized Wannier functions as a basis. The Wannier functions reflect the local symmetry of the atomic sites and the tight-binding model satisfactorily reproduces the electronic structure throughout the Brillouin zone. By constraining the hopping parameters, we have fit the isotropic conduction band effective mass, suggesting low field transport experiments can be described by the tight-binding model. By including the higher energy Ga-s(Al-s) conduction bands, the tight-binding model can also describe the parabolic to linear dispersion which suggests application to high-field electronic transport as well. Finally, by reproducing the experimental bandgap, we expect that the tight-binding model can simulate the major features in optical absorption including the primary optical transition from the O-p valence bands to the Ga-s(Al-s) conduction bands, and the optical transition from the valence band to the higher energy conduction bands. We expect the tight-binding models to aid in the description and development of electronic and optical devices utilizing bulk, nanostructured, heterostructured, and strained variants of $\beta$-Ga$_2$O$_3$, $\alpha$-Ga$_2$O$_3$, and $\alpha$-Al$_2$O$_3$.

\begin{acknowledgments}
This research project was conducted using computational resources  at the Maryland Advanced Research Computing Center (MARCC) and was supported by the National Science Foundation (Platform for the Accelerated Realization, Analysis and Discovery of Interface Materials (PARADIM)) under Cooperative Agreement No. DMR-1539918.
\end{acknowledgments}

\section*{Data Availability Statement}
The data that supports the findings of this study are available within the article and its supplementary material.

\nocite{*}
\bibliography{aipsamp}


\begin{widetext}
\clearpage
\end{widetext}

\newcommand{\beginsupplement}{%
        \setcounter{table}{0}
        \renewcommand{\thetable}{S\arabic{table}}%
        \setcounter{figure}{0}
        \renewcommand{\thefigure}{S\arabic{figure}}%
     }
     
\beginsupplement
\onecolumngrid
\begin{center}

\textbf{Supplemental Information for Tight-Binding Bandstructure of $\beta-$ and $\alpha-$ phase Ga$_2$O$_3$ and Al$_2$O$_3$}

Y. Zhang$^1$ M. Liu$^2$ D. Jena$^{1,2}$ and G. Khalsa$^2$

1. Department of Electrical and Computer Engineering and 2. Department of Materials Science and Engineering, Cornell University, Ithaca NY 14853, USA
\end{center}
\onecolumngrid

In the Supplementary Tables we include additional information for the tight-binding models of the monoclinic $\beta$-Ga$_2$O$_3$, and the rhombohedral $\alpha$-Ga$_2$O$_3$ and $\alpha$-Al$_2$O$_3$ phases. Table SI gives the structural information from density functional theory in scaled coordinates. Table SII shows the high-symmetry points in reciprocal space for the monoclinic and rhombohedral structures. Table SIII-SV give the explicit tight-binding models for each phase.

\begin{table*}[h]
\caption*{\label{tab:struct_all}TABLE SI: Structural parameters of $\beta$-Ga$_2$O$_3$, $\alpha$-Ga$_2$O$_3$, and $\alpha$-Al$_2$O$_3$. The Cartesian coordinate of a site $i$ is given by $\vec{d}_i=A_1\vec{a_1}+A_2\vec{a_2}+A_3\vec{a_3}$}
\begin{minipage}{.33\linewidth}
\begin{ruledtabular}
\begin{tabular}{ccccc}
 lattice &  \multicolumn{3}{c}{$\beta$-Ga$_2$O$_3$} \\
 vector (\AA) & $x$ & $y$ & $z$ & \\
 $\vec{a_1}$ & 6.115 & 1.620 & 0.000 & \\
 $\vec{a_2}$ & -6.115 & 1.620 & 0.000 & \\
 $\vec{a_3}$ & -1.374 & 0.000 & 5.635 & \\
 & & & & \\
 site & & & \\
 coordinate & $A_1$ & $A_2$ & $A_3$ & \\
 Ga1 & 0.090 & -0.090 & 0.795 & \\
 Ga2 & 0.910 & 0.090 & 0.205 & \\
 Ga3 & 0.659 & 0.341 & 0.314 & \\
 Ga4 & 0.341 & -0.341 & 0.686 & \\
 O1 & 0.165 & -0.165 & 0.109 & \\
 O2 & 0.835 & 0.165 & 0.891 & \\
 O3 & 0.173 & -0.173 & 0.563 & \\
 O4 & 0.827 & 0.173 & 0.437 & \\
 O5 & 0.496 & -0.496 & 0.256 & \\
 O6 & 0.504 & 0.496 & 0.743 & \\
\end{tabular}
\end{ruledtabular}
\end{minipage}%
\begin{minipage}{.33\linewidth}
\begin{ruledtabular}
\begin{tabular}{ccccc}
 lattice &  \multicolumn{3}{c}{$\alpha$-Ga$_2$O$_3$} \\
 vector (\AA) & $x$ & $y$ & $z$ & \\
 $\vec{a_1}$ & 2.491 & 1.438 & 4.478 & \\
 $\vec{a_2}$ & -2.491 & 1.438 & 4.478 & \\
 $\vec{a_3}$ & 0.000 & -2.877 & 4.478 & \\
 & & & & \\
 site & & & \\
 coordinate & $A_1$ & $A_2$ & $A_3$ & \\
Ga1 &  0.179 &  0.179 &  0.179 & \\
Ga2 &  0.678 &  0.678 & -0.322 & \\
Ga3 &  0.821 & -0.179 & -0.179 & \\
Ga4 &  0.322 &  0.322 & -0.678 & \\
O1 &  0.553 & -0.053 &  0.250 & \\
O2 &  0.941 &  0.250 & -0.441 & \\
O3 &  0.250 &  0.559 & -0.059 & \\
O4 &  1.059 & -0.250 & -0.559 & \\
O5 &  0.750 & -0.559 &  0.059 & \\
O6 &  0.441 &  0.059 & -0.250 & \\
\end{tabular}
\end{ruledtabular}
\end{minipage}%
\begin{minipage}{.33\linewidth}
\begin{ruledtabular}
\begin{tabular}{ccccc}
 lattice &  \multicolumn{3}{c}{$\alpha$-Al$_2$O$_3$} \\
 vector (\AA) & $x$ & $y$ & $z$ & \\
 $\vec{a_1}$ & 2.403 & 1.387 & 4.372 & \\
 $\vec{a_2}$ & -2.403 & 1.387 & 4.372 & \\
 $\vec{a_3}$ & 0.000 & -2.774 & 44.372 & \\
 & & & & \\
 site & & & \\
 coordinate & $A_1$ & $A_2$ & $A_3$ & \\
Al1 &  0.818 &  -0.162 & -0.177 & \\ 
Al2 &  0.671 &  0.659 & -0.311 & \\
Al3 &  0.330 &  0.318 & -0.664 & \\ 
Al4 &  0.174 &  0.182 &  0.164 & \\
O1 &  0.750 & -0.558 &  0.057 & \\
O2 &  0.942 &  0.250 & -0.442 & \\
O3 &  1.058 & -0.250 & -0.558 & \\
O4 &  0.442 &  0.058 & -0.250 & \\
O5 &  0.558 & -0.058 &  0.250 & \\
O6 &  0.250 &  0.558 & -0.058 & \\
\end{tabular}
\end{ruledtabular}
\end{minipage}%
\end{table*}

\begin{table*}[h]
\caption*{\label{tab:1BZ}TABLE SII: High-symmetry points in the first Brilloin zone of the $\beta$ and $\alpha$ phase. Coordinates are calculated by $\vec{k}=B_1\vec{b_1}+B_2\vec{b_2}+B_3\vec{b_3}$, in which $\vec{b_1}$, $\vec{b_2}$, and $\vec{b_3}$ are the reciprocal lattice vector. $\vec{b_1}$, $\vec{b_2}$, and $\vec{b_3}$ can be calculated by $\begin{pmatrix} \vec{b_1} & \vec{b_2} & \vec{b_3} \end{pmatrix} = (\begin{pmatrix} \vec{a_1} & \vec{a_2} & \vec{a_3} \end{pmatrix}^{-1})^T$}
\begin{minipage}{.4\linewidth}
\begin{ruledtabular}
\begin{tabular}{cccc}
 \multicolumn{4}{c}{$\beta$ phase} \\
 & $B_1$ & $B_2$ & $B_3$ \\
 $\Gamma$ & 0.0000 & 0.0000 & 0.0000 \\
 $C$ & 0.2662 & 0.2662 & 0.0000 \\
 $C_2$ & -0.2662 & 0.7338 & 0.0000 \\
 $Y_2$ & -0.5000 & 0.5000 & 0.0000 \\
 $M_2$ & -0.5000 & 0.5000 & 0.5000 \\
 $D$   & -0.2580 & 0.7419 & 0.5000 \\
 $D_2$ & 0.2580 & 0.2580 & 0.0000 \\
 $A$ & 0.0000 & 0.0000 & 0.5000 \\
 $L_2$ & 0.0000 & 0.5000 & 0.5000 \\
 $V_2$ & 0.0000 & 0.5000 & 0.0000
\end{tabular}
\end{ruledtabular}
\end{minipage}%
\begin{minipage}{.05\linewidth}
\begin{tabular}{c}
 \\
 \end{tabular}
\end{minipage}%
\begin{minipage}{.4\linewidth}
\begin{ruledtabular}
\begin{tabular}{cccc}
 \multicolumn{4}{c}{$\alpha$ phase} \\
 & $B_1$ & $B_2$ & $B_3$ \\
 $\Gamma$ & 0.0000 & 0.0000 & 0.0000 \\
 $T$ & 0.5000 & 0.5000 & 0.5000 \\
 $H_2$ & 0.7641 & 0.2358 & 0.5000 \\
 $H_0$ & 0.5000 & -0.2358 & 0.2358 \\
 $L$ & 0.5000 & 0.0000 & 0.0000 \\
 $S_0$ & 0.3679 & -0.3679 & 0.0000 \\
 $S_2$ & 0.6320 & 0.0000 & 0.3679 \\
 $F$ & 0.5000 & 0.0000 & 0.5000 \\
\end{tabular}
\end{ruledtabular}
\end{minipage}%
\end{table*}

\begin{table*}[ht]
\caption*{\label{tab:param_beta_GaOx}TABLE SIII: Tight-binding parameters of $\beta$-Ga$_2$O$_3$. $\vec{R}$ is written in integer multiples of lattice vectors $[ijk]$ and can be translated to Cartesian coordinates by $\vec{R}=i\vec{a_1}+j\vec{a_2}+k\vec{a_3}$, where $a_1$, $a_2$, and $a_3$ are the lattice vectors.}
\begin{minipage}{.5\linewidth}
\begin{ruledtabular}
\begin{tabular}{ccccc}
\multicolumn{2}{c}{onsite energy} & & &\\
 site $i$ & $\epsilon_i$ (eV) & & & \\
\hline
 Ga1, Ga2 & 4.95 & & & \\
 Ga3, Ga4 & 4.52 & & & \\
 all O & 0 & & & \\
 & & & & \\
\multicolumn{2}{c}{hopping parameters} & & &\\
 site $i$&site $j$&$\vec{R}$&$\tilde{t}_{ij}$ (eV)&$\phi_{ij}$\\
\hline
Ga1 s &  \multicolumn{1 }{|c}{Ga1 s} & [1,1,0] & 0.224 & $\pi$ \\
Ga2 s &  \multicolumn{1 }{|c}{Ga2 s} & [1,1,0] & 0.224 & $\pi$ \\
Ga3 s &  \multicolumn{1 }{|c}{Ga3 s} & [1,1,0] & 0.129 & 0 \\
Ga3 s &  \multicolumn{1 }{|c}{Ga4 s} & [0,0,0] & 0.108 & 0 \\
Ga3 s &  \multicolumn{1 }{|c}{Ga4 s} & [1,1,0] & 0.108 & 0 \\
Ga4 s &  \multicolumn{1 }{|c}{Ga4 s} & [1,1,0] & 0.129 & 0 \\
\hline
Ga1 s & \multicolumn{1 }{|c}{O1 px} & [0,0,1] & 0.652 & 0 \\
Ga1 s & \multicolumn{1 }{|c}{O1 pz} & [0,0,1] & 3.467 & 0 \\ \cline{2-5}
Ga1 s & \multicolumn{1 }{|c}{O3 px} & [0,0,0] & 2.592 & 0 \\
Ga1 s & \multicolumn{1 }{|c}{O3 pz} & [0,0,0] & 2.592 & $\pi$ \\ \cline{2-5}
Ga1 s & \multicolumn{1 }{|c}{O6 px} & [0,0,0] & 1.748 & $\pi$ \\
Ga1 s & \multicolumn{1 }{|c}{O6 px} & [-1,-1,0] & 1.748 & $\pi$ \\
Ga1 s & \multicolumn{1 }{|c}{O6 py} & [0,0,0] & 3.029 & 0 \\
Ga1 s & \multicolumn{1 }{|c}{O6 py} & [-1,-1,0] & 3.029 & $\pi$ \\
Ga1 s & \multicolumn{1 }{|c}{O6 pz} & [0,0,0] & 0.440 & $\pi$ \\
Ga1 s & \multicolumn{1 }{|c}{O6 pz} & [-1,-1,0] & 0.440 & $\pi$ \\
\hline
Ga2 s & \multicolumn{1 }{|c}{O2 px} & [0,0,-1] & 0.652 & $\pi$ \\
Ga2 s & \multicolumn{1 }{|c}{O2 pz} & [0,0,-1] & 3.467 & $\pi$ \\ \cline{2-5}
Ga2 s & \multicolumn{1 }{|c}{O4 px} & [0,0,0] & 2.596 & $\pi$ \\
Ga2 s & \multicolumn{1 }{|c}{O4 pz} & [0,0,0] & 2.596 & 0 \\ \cline{2-5}
Ga2 s & \multicolumn{1 }{|c}{O5 px} & [0,0,0] & 1.750 & 0 \\
Ga2 s & \multicolumn{1 }{|c}{O5 px} & [1,1,0] & 1.750 & 0 \\
Ga2 s & \multicolumn{1 }{|c}{O5 py} & [0,0,0] & 3.030 & $\pi$ \\
Ga2 s & \multicolumn{1 }{|c}{O5 py} & [1,1,0] & 3.030 & 0 \\
Ga2 s & \multicolumn{1 }{|c}{O5 pz} & [0,0,0] & 0.443 & 0 \\
Ga2 s & \multicolumn{1 }{|c}{O5 pz} & [1,1,0] & 0.443 & 0 \\
\end{tabular}
\end{ruledtabular}
\end{minipage}%
\begin{minipage}{.02\linewidth}
\begin{tabular}{c}
 \\
 \end{tabular}
\end{minipage}%
\begin{minipage}{.5\linewidth}
\begin{ruledtabular}
\begin{tabular}{ccccc}
\multicolumn{3}{c}{hopping parameters continued} & &\\
 site $i$&site $j$&$\vec{R}$&$\tilde{t}_{ij}$ (eV)&$\phi_{ij}$\\
 \hline
Ga3 s & \multicolumn{1 }{|c}{O1 px} & [0,0,0] & 0.355 & 0 \\
Ga3 s & \multicolumn{1 }{|c}{O1 px} & [1,1,0] & 0.355 & 0 \\
Ga3 s & \multicolumn{1 }{|c}{O1 py} & [0,0,0] & 2.484 & $\pi$ \\
Ga3 s & \multicolumn{1 }{|c}{O1 py} & [1,1,0] & 2.484 & 0 \\
Ga3 s & \multicolumn{1 }{|c}{O1 pz} & [0,0,0] & 1.654 & $\pi$ \\
Ga3 s & \multicolumn{1 }{|c}{O1 pz} & [1,1,0] & 1.654 & $\pi$ \\ \cline{2-5}
Ga3 s & \multicolumn{1 }{|c}{O3 py} & [0,0,0] & 2.162 & $\pi$ \\
Ga3 s & \multicolumn{1 }{|c}{O3 py} & [1,1,0] & 2.162 & 0 \\
Ga3 s & \multicolumn{1 }{|c}{O3 pz} & [0,0,0] & 1.673 & 0 \\
Ga3 s & \multicolumn{1 }{|c}{O3 pz} & [1,1,0] & 1.673 & 0 \\ \cline{2-5}
Ga3 s & \multicolumn{1 }{|c}{O4 px} & [0,0,0] & 2.878 & 0 \\
Ga3 s & \multicolumn{1 }{|c}{O4 pz} & [0,0,0] & 1.102 & 0 \\ \cline{2-5}
Ga3 s & \multicolumn{1 }{|c}{O5 px} & [0,1,0] & 2.965 & $\pi$ \\
Ga3 s & \multicolumn{1 }{|c}{O5 pz} & [0,1,0] & 0.622 & $\pi$ \\
\hline
Ga4 s & \multicolumn{1 }{|c}{O2 px} & [0,0,0] & 0.352 & $\pi$ \\
Ga4 s & \multicolumn{1 }{|c}{O2 px} & [-1,-1,0] & 0.352 & $\pi$ \\
Ga4 s & \multicolumn{1 }{|c}{O2 py} & [0,0,0] & 2.484 & 0 \\
Ga4 s & \multicolumn{1 }{|c}{O2 py} & [-1,-1,0] & 2.484 & $\pi$ \\
Ga4 s & \multicolumn{1 }{|c}{O2 pz} & [0,0,0] & 1.657 & 0 \\
Ga4 s & \multicolumn{1 }{|c}{O2 pz} & [-1,-1,0] & 1.657 & 0 \\ \cline{2-5}
Ga4 s & \multicolumn{1 }{|c}{O3 px} & [0,0,0] & 2.878 & $\pi$ \\
Ga4 s & \multicolumn{1 }{|c}{O3 pz} & [0,0,0] & 1.102 & $\pi$ \\ \cline{2-5}
Ga4 s & \multicolumn{1 }{|c}{O4 py} & [0,0,0] & 2.164 & 0 \\
Ga4 s & \multicolumn{1 }{|c}{O4 py} & [-1,-1,0] & 2.164 & $\pi$ \\
Ga4 s & \multicolumn{1 }{|c}{O4 pz} & [0,0,0] & 1.663 & $\pi$ \\
Ga4 s & \multicolumn{1 }{|c}{O4 pz} & [-1,-1,0] & 1.663 & $\pi$ \\ \cline{2-5}
Ga4 s & \multicolumn{1 }{|c}{O6 px} & [0,-1,0] & 2.965 & 0 \\
Ga4 s & \multicolumn{1 }{|c}{O6 pz} & [0,-1,0] & 0.622 & 0 \\
\end{tabular}
\end{ruledtabular}
\end{minipage}%
\end{table*}

\begin{table*}[ht]
\caption*{\label{tab:param_alpha_GaOx}TABLE SIV: Tight-binding parameters of $\alpha$-Ga$_2$O$_3$. $\vec{R}$ is written in integer multiples of lattice vectors $[ijk]$ and can be translated to Cartesian coordinates by $\vec{R}=i\vec{a_1}+j\vec{a_2}+k\vec{a_3}$, where $a_1$, $a_2$, and $a_3$ are the lattice vectors.}
\begin{minipage}{.5\linewidth}
\begin{ruledtabular}
\begin{tabular}{ccccc}
\multicolumn{2}{c}{onsite energy} & & &\\
 site $i$ & $\epsilon_i$ (eV) & & & \\
\hline
 all Ga & -5.48 & & & \\
 all O & -10.5 & & & \\
 & & & & \\
\multicolumn{2}{c}{hopping parameters} & & &\\
 site $i$&site $j$&$\vec{R}$&$\tilde{t}_{ij}$&$\phi_{ij}$\\
\hline
Ga1 s & \multicolumn{1 }{|c}{Ga3 s} & [0,0,0] & 0.012 & 0 \\
Ga1 s & \multicolumn{1 }{|c}{Ga3 s} & [-1,1,0] & 0.013 & 0 \\
Ga1 s & \multicolumn{1 }{|c}{Ga3 s} & [-1,0,1] & 0.013 & 0 \\
Ga1 s & \multicolumn{1 }{|c}{Ga4 s} & [0,0,1] & 0.014 & $\pi$ \\
Ga1 s & \multicolumn{1 }{|c}{Ga4 s} & [0,-1,1] & 0.010 & $\pi$ \\
Ga1 s & \multicolumn{1 }{|c}{Ga4 s} & [-1,0,1] & 0.010 & $\pi$ \\
Ga2 s & \multicolumn{1 }{|c}{Ga3 s} & [0,0,0] & 0.010 & $\pi$ \\
Ga2 s & \multicolumn{1 }{|c}{Ga3 s} & [0,1,0] & 0.014 & $\pi$ \\
Ga2 s & \multicolumn{1 }{|c}{Ga3 s} & [-1,1,0] & 0.010 & $\pi$ \\
Ga2 s & \multicolumn{1 }{|c}{Ga4 s} & [0,0,1] & 0.013 & 0 \\
Ga2 s & \multicolumn{1 }{|c}{Ga4 s} & [0,1,0] & 0.013 & 0 \\
Ga2 s & \multicolumn{1 }{|c}{Ga4 s} & [1,0,0] & 0.013 & 0 \\
\hline
Ga1 s & \multicolumn{1 }{|c}{O1 px} & [0,0,0] & 1.983 & 0 \\
Ga1 s & \multicolumn{1 }{|c}{O1 py} & [0,0,0] & 0.105 & 0 \\
Ga1 s & \multicolumn{1 }{|c}{O1 pz} & [0,0,0] & 1.860 & 0 \\ \cline{2-5}
Ga1 s & \multicolumn{1 }{|c}{O2 px} & [-1,0,1] & 0.910 & $\pi$ \\
Ga1 s & \multicolumn{1 }{|c}{O2 py} & [-1,0,1] & 1.766 & $\pi$ \\
Ga1 s & \multicolumn{1 }{|c}{O2 pz} & [-1,0,1] & 1.864 & 0 \\ \cline{2-5}
Ga1 s & \multicolumn{1 }{|c}{O3 px} & [0,0,0] & 1.076 & $\pi$ \\
Ga1 s & \multicolumn{1 }{|c}{O3 py} & [0,0,0] & 1.673 & 0 \\
Ga1 s & \multicolumn{1 }{|c}{O3 pz} & [0,0,0] & 1.867 & 0 \\ \cline{2-5}
Ga1 s & \multicolumn{1 }{|c}{O4 px} & [-1,0,1] & 1.151 & 0 \\
Ga1 s & \multicolumn{1 }{|c}{O4 py} & [-1,0,1] & 2.402 & $\pi$ \\
Ga1 s & \multicolumn{1 }{|c}{O4 pz} & [-1,0,1] & 1.248 & $\pi$ \\ \cline{2-5}
Ga1 s & \multicolumn{1 }{|c}{O5 px} & [-1,1,0] & 2.654 & $\pi$ \\
Ga1 s & \multicolumn{1 }{|c}{O5 py} & [-1,1,0] & 0.216 & 0 \\
Ga1 s & \multicolumn{1 }{|c}{O5 pz} & [-1,1,0] & 1.246 & $\pi$ \\ \cline{2-5}
Ga1 s & \multicolumn{1 }{|c}{O6 px} & [0,0,0] & 1.509 & 0 \\
Ga1 s & \multicolumn{1 }{|c}{O6 py} & [0,0,0] & 2.197 & 0 \\
Ga1 s & \multicolumn{1 }{|c}{O6 pz} & [0,0,0] & 1.242 & $\pi$ \\
\hline
Ga2 s & \multicolumn{1 }{|c}{O1 px} & [0,1,-1] & 1.511 & $\pi$ \\
Ga2 s & \multicolumn{1 }{|c}{O1 py} & [0,1,-1] & 2.198 & 0 \\
Ga2 s & \multicolumn{1 }{|c}{O1 pz} & [0,1,-1] & 1.249 & $\pi$ \\ \cline{2-5}
Ga2 s & \multicolumn{1 }{|c}{O2 px} & [0,0,0] & 2.656 & 0 \\
Ga2 s & \multicolumn{1 }{|c}{O2 py} & [0,0,0] & 0.216 & 0 \\
Ga2 s & \multicolumn{1 }{|c}{O2 pz} & [0,0,0] & 1.242 & $\pi$ \\ \cline{2-5}
Ga2 s & \multicolumn{1 }{|c}{O3 px} & [0,0,0] & 1.142 & $\pi$ \\
Ga2 s & \multicolumn{1 }{|c}{O3 py} & [0,0,0] & 2.404 & $\pi$ \\
Ga2 s & \multicolumn{1 }{|c}{O3 pz} & [0,0,0] & 1.245 & $\pi$ \\ \cline{2-5}
Ga2 s & \multicolumn{1 }{|c}{O4 px} & [0,1,0] & 1.074 & 0 \\
Ga2 s & \multicolumn{1 }{|c}{O4 py} & [0,1,0] & 1.664 & 0 \\
Ga2 s & \multicolumn{1 }{|c}{O4 pz} & [0,1,0] & 1.866 & 0 \\ \cline{2-5}
Ga2 s & \multicolumn{1 }{|c}{O5 px} & [0,1,0] & 0.904 & 0 \\
Ga2 s & \multicolumn{1 }{|c}{O5 py} & [0,1,0] & 1.759 & $\pi$ \\
Ga2 s & \multicolumn{1 }{|c}{O5 pz} & [0,1,0] & 1.872 & 0 \\ \cline{2-5}
Ga2 s & \multicolumn{1 }{|c}{O6 px} & [0,1,0] & 1.980 & $\pi$ \\
Ga2 s & \multicolumn{1 }{|c}{O6 py} & [0,1,0] & 0.096 & 0 \\
Ga2 s & \multicolumn{1 }{|c}{O6 pz} & [0,1,0] & 1.870 & 0 \\
\end{tabular}
\end{ruledtabular}
\end{minipage}%
\begin{minipage}{.02\linewidth}
\begin{tabular}{c}
 \\
 \end{tabular}
\end{minipage}%
\begin{minipage}{.5\linewidth}
\begin{ruledtabular}
\begin{tabular}{ccccc}
\multicolumn{3}{c}{hopping parameters continued} & &\\
 site $i$&site $j$&$\vec{R}$&$\tilde{t}_{ij}$ (eV)&$\phi_{ij}$\\
 \hline
Ga3 s & \multicolumn{1 }{|c}{O1 px} & [0,0,0] & 1.508 & $\pi$ \\
Ga3 s & \multicolumn{1 }{|c}{O1 py} & [0,0,0] & 2.191 & $\pi$ \\
Ga3 s & \multicolumn{1 }{|c}{O1 pz} & [0,0,0] & 1.252 & 0 \\ \cline{2-5}
Ga3 s & \multicolumn{1 }{|c}{O2 px} & [0,0,0] & 1.151 & $\pi$ \\
Ga3 s & \multicolumn{1 }{|c}{O2 py} & [0,0,0] & 2.402 & 0 \\
Ga3 s & \multicolumn{1 }{|c}{O2 pz} & [0,0,0] & 1.251 & 0 \\ \cline{2-5}
Ga3 s & \multicolumn{1 }{|c}{O3 px} & [1,-1,0] & 2.656 & 0 \\
Ga3 s & \multicolumn{1 }{|c}{O3 py} & [1,-1,0] & 0.199 & $\pi$ \\
Ga3 s & \multicolumn{1 }{|c}{O3 pz} & [1,-1,0] & 1.253 & 0 \\ \cline{2-5}
Ga3 s & \multicolumn{1 }{|c}{O4 px} & [0,0,0] & 0.910 & 0 \\
Ga3 s & \multicolumn{1 }{|c}{O4 py} & [0,0,0] & 1.779 & 0 \\
Ga3 s & \multicolumn{1 }{|c}{O4 pz} & [0,0,0] & 1.851 & $\pi$ \\ \cline{2-5}
Ga3 s & \multicolumn{1 }{|c}{O5 px} & [0,0,0] & 1.084 & 0 \\
Ga3 s & \multicolumn{1 }{|c}{O5 py} & [0,0,0] & 1.681 & $\pi$ \\
Ga3 s & \multicolumn{1 }{|c}{O5 pz} & [0,0,0] & 1.853 & $\pi$ \\ \cline{2-5}
Ga3 s & \multicolumn{1 }{|c}{O6 px} & [0,0,0] & 1.995 & $\pi$ \\
Ga3 s & \multicolumn{1 }{|c}{O6 py} & [0,0,0] & 0.098 & $\pi$ \\
Ga3 s & \multicolumn{1 }{|c}{O6 pz} & [0,0,0] & 1.856 & $\pi$ \\
\hline
Ga4 s & \multicolumn{1 }{|c}{O1 px} & [0,0,-1] & 1.989 & 0 \\
Ga4 s & \multicolumn{1 }{|c}{O1 py} & [0,0,-1] & 0.098 & $\pi$ \\
Ga4 s & \multicolumn{1 }{|c}{O1 pz} & [0,0,-1] & 1.859 & $\pi$ \\ \cline{2-5}
Ga4 s & \multicolumn{1 }{|c}{O2 px} & [-1,0,0] & 1.077 & $\pi$ \\
Ga4 s & \multicolumn{1 }{|c}{O2 py} & [-1,0,0] & 1.677 & $\pi$ \\
Ga4 s & \multicolumn{1 }{|c}{O2 pz} & [-1,0,0] & 1.862 & $\pi$ \\ \cline{2-5}
Ga4 s & \multicolumn{1 }{|c}{O3 px} & [0,0,-1] & 0.908 & $\pi$ \\
Ga4 s & \multicolumn{1 }{|c}{O3 py} & [0,0,-1] & 1.775 & 0 \\
Ga4 s & \multicolumn{1 }{|c}{O3 pz} & [0,0,-1] & 1.856 & $\pi$ \\ \cline{2-5}
Ga4 s & \multicolumn{1 }{|c}{O4 px} & [-1,1,0] & 2.653 & $\pi$ \\
Ga4 s & \multicolumn{1 }{|c}{O4 py} & [-1,1,0] & 0.213 & $\pi$ \\
Ga4 s & \multicolumn{1 }{|c}{O4 pz} & [-1,1,0] & 1.253 & 0 \\ \cline{2-5}
Ga4 s & \multicolumn{1 }{|c}{O5 px} & [0,1,-1] & 1.151 & 0 \\
Ga4 s & \multicolumn{1 }{|c}{O5 py} & [0,1,-1] & 2.405 & 0 \\
Ga4 s & \multicolumn{1 }{|c}{O5 pz} & [0,1,-1] & 1.249 & 0 \\ \cline{2-5}
Ga4 s & \multicolumn{1 }{|c}{O6 px} & [0,0,0] & 1.510 & 0 \\
Ga4 s & \multicolumn{1 }{|c}{O6 py} & [0,0,0] & 2.196 & $\pi$ \\
Ga4 s & \multicolumn{1 }{|c}{O6 pz} & [0,0,0] & 1.254 & 0 \\
\end{tabular}
\end{ruledtabular}
\end{minipage}%
\end{table*}

\begin{table*}[ht]
\caption*{\label{tab:param_alpha_AlOx}TABLE SV: Tight-binding parameters of $\alpha$-Al$_2$O$_3$. $\vec{R}$ is written in integer multiples of lattice vectors $[ijk]$ and can be translated to Cartesian coordinates by $\vec{R}=i\vec{a_1}+j\vec{a_2}+k\vec{a_3}$, where $a_1$, $a_2$, and $a_3$ are the lattice vectors.}
\begin{minipage}{.5\linewidth}
\begin{ruledtabular}
\begin{tabular}{ccccc}
\multicolumn{2}{c}{onsite energy (eV)} & & &\\
 all Al & 7.00 & & & \\
 all O & 0.00 & & & \\
 & & & & \\
\multicolumn{2}{c}{hopping parameters} & & &\\
 site $i$&site $j$&$\vec{R}$&$\tilde{t}_{ij}$&$\phi_{ij}$\\
\hline
Al1 s & \multicolumn{1 }{|c}{Al2 s} & [0,-1,0] & 1.019 & $\pi$ \\
Al1 s & \multicolumn{1 }{|c}{Al3 s} & [0,0,0] & 0.429 & 0 \\
Al1 s & \multicolumn{1 }{|c}{Al3 s} & [0,0,1] & 0.389 & 0 \\
Al1 s & \multicolumn{1 }{|c}{Al3 s} & [1,0,0] & 0.405 & 0 \\
Al2 s & \multicolumn{1 }{|c}{Al4 s} & [0,0,0] & 0.406 & 0 \\
Al2 s & \multicolumn{1 }{|c}{Al4 s} & [1,0,0] & 0.415 & 0 \\
Al3 s & \multicolumn{1 }{|c}{Al4 s} & [0,0,-1] & 1.024 & $\pi$ \\
\hline
Al1 s & \multicolumn{1 }{|c}{O1 px} & [0,0,0] & 1.592 & 0 \\
Al1 s & \multicolumn{1 }{|c}{O1 py} & [0,0,0] & 2.614 & $\pi$ \\
Al1 s & \multicolumn{1 }{|c}{O1 pz} & [0,0,0] & 2.646 & $\pi$ \\ \cline{2-5}
Al1 s & \multicolumn{1 }{|c}{O2 px} & [0,0,0] & 1.008 & $\pi$ \\
Al1 s & \multicolumn{1 }{|c}{O2 py} & [0,0,0] & 2.321 & 0 \\
Al1 s & \multicolumn{1 }{|c}{O2 pz} & [0,0,0] & 1.738 & 0 \\ \cline{2-5}
Al1 s & \multicolumn{1 }{|c}{O3 px} & [0,0,0] & 1.553 & 0 \\
Al1 s & \multicolumn{1 }{|c}{O3 py} & [0,0,0] & 2.708 & 0 \\
Al1 s & \multicolumn{1 }{|c}{O3 pz} & [0,0,0] & 2.685 & $\pi$ \\ \cline{2-5}
Al1 s & \multicolumn{1 }{|c}{O4 px} & [0,0,0] & 3.249 & $\pi$ \\
Al1 s & \multicolumn{1 }{|c}{O4 pz} & [0,0,0] & 2.871 & $\pi$ \\ \cline{2-5}
Al1 s & \multicolumn{1 }{|c}{O5 px} & [0,0,0] & 1.392 & $\pi$ \\
Al1 s & \multicolumn{1 }{|c}{O5 py} & [0,0,0] & 2.058 & $\pi$ \\
Al1 s & \multicolumn{1 }{|c}{O5 pz} & [0,0,0] & 1.717 & 0 \\ \cline{2-5}
Al1 s & \multicolumn{1 }{|c}{O6 px} & [1,-1,0] & 2.352 & 0 \\
Al1 s & \multicolumn{1 }{|c}{O6 py} & [1,-1,0] & 0.250 & $\pi$ \\
Al1 s & \multicolumn{1 }{|c}{O6 pz} & [1,-1,0] & 1.597 & 0 \\
\hline
Al2 s & \multicolumn{1 }{|c}{O1 px} & [0,1,0] & 1.578 & 0 \\
Al2 s & \multicolumn{1 }{|c}{O1 py} & [0,1,0] & 2.783 & $\pi$ \\
Al2 s & \multicolumn{1 }{|c}{O1 pz} & [0,1,0] & 2.994 & 0 \\ \cline{2-5}
Al2 s & \multicolumn{1 }{|c}{O2 px} & [0,0,0] & 2.511 & 0 \\
Al2 s & \multicolumn{1 }{|c}{O2 py} & [0,0,0] & 0.365 & 0 \\
Al2 s & \multicolumn{1 }{|c}{O2 pz} & [0,0,0] & 1.704 & $\pi$ \\ \cline{2-5}
Al2 s & \multicolumn{1 }{|c}{O3 px} & [0,1,0] & 1.473 & 0 \\
Al2 s & \multicolumn{1 }{|c}{O3 py} & [0,1,0] & 2.567 & 0 \\
Al2 s & \multicolumn{1 }{|c}{O3 pz} & [0,1,0] & 2.657 & 0 \\ \cline{2-5}
Al2 s & \multicolumn{1 }{|c}{O4 px} & [0,1,0] & 2.998 & $\pi$ \\
Al2 s & \multicolumn{1 }{|c}{O4 pz} & [0,1,0] & 2.722 & 0 \\ \cline{2-5}
Al2 s & \multicolumn{1 }{|c}{O5 px} & [0,1,-1] & 1.381 & $\pi$ \\
Al2 s & \multicolumn{1 }{|c}{O5 py} & [0,1,-1] & 1.901 & 0 \\
Al2 s & \multicolumn{1 }{|c}{O5 pz} & [0,1,-1] & 1.554 & $\pi$ \\ \cline{2-5}
Al2 s & \multicolumn{1 }{|c}{O6 px} & [0,0,0] & 1.144 & $\pi$ \\
Al2 s & \multicolumn{1 }{|c}{O6 py} & [0,0,0] & 2.297 & $\pi$ \\
Al2 s & \multicolumn{1 }{|c}{O6 pz} & [0,0,0] & 1.769 & $\pi$ \\
\end{tabular}
\end{ruledtabular}
\end{minipage}%
\begin{minipage}{.02\linewidth}
\begin{tabular}{c}
 \\
 \end{tabular}
\end{minipage}%
\begin{minipage}{.5\linewidth}
\begin{ruledtabular}
\begin{tabular}{ccccc}
\multicolumn{3}{c}{hopping parameters continued} & &\\
 site $i$&site $j$&$\vec{R}$&$\tilde{t}_{ij}$ (eV)&$\phi_{ij}$\\
 \hline
Al3 s & \multicolumn{1 }{|c}{O1 px} & [0,1,-1] & 0.924 & 0 \\
Al3 s & \multicolumn{1 }{|c}{O1 py} & [0,1,-1] & 2.264 & 0 \\
Al3 s & \multicolumn{1 }{|c}{O1 pz} & [0,1,-1] & 1.620 & 0 \\ \cline{2-5}
Al3 s & \multicolumn{1 }{|c}{O2 px} & [-1,0,0] & 1.644 & $\pi$ \\
Al3 s & \multicolumn{1 }{|c}{O2 py} & [-1,0,0] & 2.699 & $\pi$ \\
Al3 s & \multicolumn{1 }{|c}{O2 pz} & [-1,0,0] & 2.806 & $\pi$ \\ \cline{2-5}
Al3 s & \multicolumn{1 }{|c}{O3 px} & [-1,1,0] & 2.385 & $\pi$ \\
Al3 s & \multicolumn{1 }{|c}{O3 py} & [-1,1,0] & 0.189 & $\pi$ \\
Al3 s & \multicolumn{1 }{|c}{O3 pz} & [-1,1,0] & 1.636 & 0 \\ \cline{2-5}
Al3 s & \multicolumn{1 }{|c}{O4 px} & [0,0,0] & 1.476 & 0 \\
Al3 s & \multicolumn{1 }{|c}{O4 py} & [0,0,0] & 2.108 & $\pi$ \\
Al3 s & \multicolumn{1 }{|c}{O4 pz} & [0,0,0] & 1.798 & 0 \\ \cline{2-5}
Al3 s & \multicolumn{1 }{|c}{O5 px} & [0,0,-1] & 3.176 & 0 \\
Al3 s & \multicolumn{1 }{|c}{O5 pz} & [0,0,-1] & 2.857 & $\pi$ \\ \cline{2-5}
Al3 s & \multicolumn{1 }{|c}{O6 px} & [0,0,-1] & 1.451 & $\pi$ \\
Al3 s & \multicolumn{1 }{|c}{O6 py} & [0,0,-1] & 2.631 & 0 \\
Al3 s & \multicolumn{1 }{|c}{O6 pz} & [0,0,-1] & 2.596 & $\pi$ \\
\hline
Al4 s & \multicolumn{1 }{|c}{O1 px} & [-1,1,0] & 2.510 & $\pi$ \\
Al4 s & \multicolumn{1 }{|c}{O1 py} & [-1,1,0] & 0.197 & 0 \\
Al4 s & \multicolumn{1 }{|c}{O1 pz} & [-1,1,0] & 1.727 & $\pi$ \\ \cline{2-5}
Al4 s & \multicolumn{1 }{|c}{O2 px} & [-1,0,1] & 1.468 & $\pi$ \\
Al4 s & \multicolumn{1 }{|c}{O2 py} & [-1,0,1] & 2.660 & $\pi$ \\
Al4 s & \multicolumn{1 }{|c}{O2 pz} & [-1,0,1] & 2.715 & 0 \\ \cline{2-5}
Al4 s & \multicolumn{1 }{|c}{O3 px} & [-1,0,1] & 0.974 & 0 \\
Al4 s & \multicolumn{1 }{|c}{O3 py} & [-1,0,1] & 2.175 & $\pi$ \\
Al4 s & \multicolumn{1 }{|c}{O3 pz} & [-1,0,1] & 1.606 & $\pi$ \\ \cline{2-5}
Al4 s & \multicolumn{1 }{|c}{O4 px} & [0,0,0] & 1.535 & 0 \\
Al4 s & \multicolumn{1 }{|c}{O4 py} & [0,0,0] & 2.022 & 0 \\
Al4 s & \multicolumn{1 }{|c}{O4 pz} & [0,0,0] & 1.687 & $\pi$ \\ \cline{2-5}
Al4 s & \multicolumn{1 }{|c}{O5 px} & [0,0,0] & 3.038 & 0 \\
Al4 s & \multicolumn{1 }{|c}{O5 pz} & [0,0,0] & 2.702 & 0 \\ \cline{2-5}
Al4 s & \multicolumn{1 }{|c}{O6 px} & [0,0,0] & 1.619 & $\pi$ \\
Al4 s & \multicolumn{1 }{|c}{O6 py} & [0,0,0] & 2.746 & 0 \\
Al4 s & \multicolumn{1 }{|c}{O6 pz} & [0,0,0] & 2.903 & 0 \\
\end{tabular}
\end{ruledtabular}
\end{minipage}%
\end{table*}

\end{document}